\documentclass[preprint,showpacs,preprintnumbers,amsmath,amssymb]{revtex4}

\usepackage{graphicx}
\usepackage{dcolumn}
\usepackage{bm}

\def\be{\begin{equation}}
\def\ee{\end{equation}}
\def\bea{\begin{eqnarray}}
\def\eea{\end{eqnarray}}
\def\pt{\partial}
\def\ffi{\varphi}

\def\dt{\delta}
\def\al{\alpha}
\def\gm{\gamma}

\def\Th{\Theta}
\def\eps{\varepsilon}
\def\Dt{\Delta}

\def\la{\lambda}

\def\om{\omega}

\def\O{\mbox{O}}

\def\mod{\mbox{mod}}

\def\dd{\mbox{d}}

\def\calP{{\cal P}}
\def\calH{{\cal H}}
\def\calF{{\cal F}}
\def\calS{{\cal S}}
\def\calJ{{\cal J}}
\def\tcalF{\tilde{\cal F}}

\begin{document}

\title{Capture into resonance in dynamics of a classical hydrogen atom in an oscillating electric
field }

\author{Anatoly Neishtadt and Alexei Vasiliev}
\email{aneishta@iki.rssi.ru, valex@iki.rssi.ru}

\affiliation{Space Research Institute,\\
Profsoyuznaya str. 84/32, 117997 Moscow, Russia}


\begin{abstract}

We consider a classical hydrogen atom in a linearly polarized electric field of slow changing
frequency. When the system passes through a resonance between the driving frequency and the
Keplerian frequency of the electron's motion, a capture into the resonance can occur. We study this
phenomenon in the case of 2:1 resonance and show that the capture results in growth of the
eccentricity of the electron's orbit. The capture probability for various initial values of the
eccentricity is defined and calculated.

\end{abstract}


\maketitle

\section{ Introduction}\label{intro}

Dynamics of highly excited (Rydberg) atoms in microwave fields has been a subject of extensive
research during the last thirty years. After experiments of Bayfield and Koch \cite{BK} and
theoretical work of Leopold and Percival \cite{LP}, it was realised that certain essential
properties of the dynamics of Rydberg hydrogen atoms can be described in the frames of classical
approach.

One of the classical ideas in this area is to control the Keplerian motion of the electron using
its resonant interaction with a wave of slowly changing frequency. Dynamical problems of this kind
were studied in \cite{MF} for a 1-D model and \cite{GF} for a 3-D model. In particular, in the
latter work an hydrogen atom in a linearly polarized electric field of slowly decreasing frequency
was considered. It was shown that at a passage through 2:1 resonance (i.e., when the driving
frequency is twice as large as the Keplerian frequency) the system with initially zero eccentricity
of the electron's orbit is captured into the resonance. In the captured state, the electron's
Keplerian frequency varies in such a way that the resonant condition is approximately satisfied. In
this motion the orbit's eccentricity grows, which may result in ionization of the atom.

In the present work we also consider a 3-D hydrogen atom in a linearly polarized electrostatic
field of slowly changing frequency. We study behaviour of the system near 2:1 resonance using
methods of the theory of resonant phenomena, developed in \cite{N75} - \cite{N99} (see also
\cite{AKN}). These methods were previously used in studies of various physical problems including
surfatron acceleration of charged particles in magnetic field and electromagnetic wave
\cite{surfatron}, slowly perturbed billiards \cite{bill1}, \cite{bill2}, and classical dynamics in
a molecular hydrogen ion \cite{It2003}. In the present paper, we show that capture into the
resonance necessarily occurs not only in the case of zero initial eccentricity but also if the
initial eccentricity is not zero but small enough. Moreover, at larger values of the initial
eccentricity the capture is also possible. Following the general approach, capture into the
resonance in this case can be considered as a probabilistic phenomenon. We define and evaluate its
probability. The obtained results can be used to broaden the applicability of the resonant control
methods for Rydberg atoms.

The paper is organized as follows. In Section 2, we use standard techniques of classical celestial
mechanics and the theory of resonant phenomena to reduce the equations of motion near the resonance
to the standard form. We consider two different cases: one of small eccentricity and the other of
eccentricity of order 1. In Section 3, we study the case of small eccentricity. We apply relevant
results of \cite{N75} and find the region of so called "automatic capture" into the resonance at
small eccentricities and calculate probability of capture at larger values of initial eccentricity.
Section 4 is devoted to the capture phenomenon at values of eccentricity of order 1. We calculate
the capture probability in this case too. In the both cases, the capture significantly changes the
eccentricity of the electron's orbit and may lead to ionization. In Conclusions, we summarize the
results.

\section{ Equations of motion near the 2:1 resonance}\label{equations}

We study dynamics of a classical electron in a hydrogen atom perturbed by a harmonically
oscillating electric field of small amplitude $\mu$, linearly polarized along the $Z$-axis. This
system is described with Hamiltonian
\be
\calH = \calH_0 + Z\mu \cos \Psi.
\label{2.1}
\ee
Here $\calH_0$ is the unperturbed Hamiltonian of motion in the Coulomb field and $\Psi$ is the
perturbation phase. Introduce the perturbation frequency $\om = d\Psi/dt$. We assume that $\om =
\om(\eps t),\, 0<\eps \ll 1$, i.e. that this frequency slowly changes with time. For brevity, we
put the electron mass and charge to 1, and use dimensionless variables.

\begin{figure*}
\includegraphics{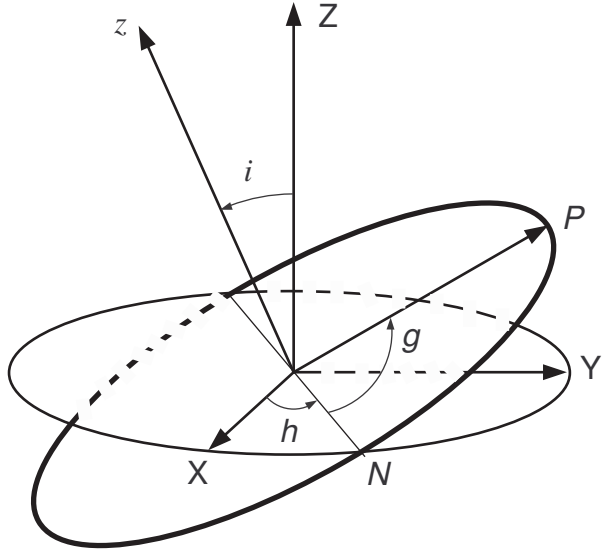}
\caption{\label{orbit} The Keplerian ellipse is shown with the bold line. The rest system of
coordinates is $(XYZ)$, the $z$-axis is orthogonal to the plane of the orbit. The periapsis and the
ascending node are denoted with P and N accordingly. }
\end{figure*}

The unperturbed trajectory of the electron is an ellipse with eccentricity $e$, semimajor axis $a$,
and inclination $i$. It is a well-known fact from celestial mechanics that the so-called Delaunay
elements $L,G,H,l,g,h$ provide a set of canonical variables for the system under consideration
(see, e.g. \cite{BrCl}). The Delaunay elements can be defined as

\be
L = \sqrt{a},\;\; G = \sqrt{a(1-e^2)},\;\; H = \sqrt{a(1-e^2)}\cos i;
\label{2.2}
\ee
$l$ is the mean anomaly, $g$ is the argument of the periapsis, and $h$ is the longitude
of the ascending node (see Figure \ref{orbit}).

In these variables, Hamiltonian (\ref{2.1}) takes the form (see \cite{GF}):
\bea
\calH &=& \calH_0 + \mu\calH_1, \;\; \calH_0 = -\frac{1}{2L^2},
\nonumber \\
\calH_1 &=& \sin i \sum^{\infty}_{k=1} \left[\frac{a_k +b_k}{4} \left(\sin(kl+g-\Psi)
+\sin(kl+g+\Psi)\right)\right. +
\label{2.3} \\
& &\left.\frac{b_k -a_k}{4} \left(\sin(kl-g-\Psi) +\sin(kl-g+\Psi)\right)\right], \nonumber
\eea
where
\be
a_k = \frac{2a}{k} \calJ '_k (k e),\; b_k = \frac{2a\sqrt{1-e^2}}{k e} \calJ_k(k e).
\label{2.3a}
\ee
Here $\calJ_k(\cdot)$ is the Bessel function of integer order $k$, and $\calJ'_k(\cdot)$ is its
derivative.

In order to avoid possible singularities at $e=0$, we make a canonical transformation of variables
$(L,G,H;l,g,h) \mapsto (P_3,P_2,P_1;Q_3,Q_2,Q_1)$ defined with generating function $P_3(l+g+h) +
P_2(g+h) + P_3h$. The new canonical variables [called Poincar\'e elements of the first kind] are
expressed in terms of the old ones as follows:

\be
\begin{array}{cc}
P_3 = L, & Q_3 = l+g+h,
\\
P_2 = G-L, & Q_2 = g+h,
\\
P_1 = H-G, & Q_1 = h.
\end{array}
\label{2.4}
\ee

As the perturbation frequency slowly varies with time, the system passes through resonances with
the unperturbed Keplerian frequency $\dot l$. Near a resonance, certain terms in expression
(\ref{2.3}) for $\calH_1$ are changing very slowly. Consider a passage through the 2:1 resonance.
In this case, after averaging over fast oscillating terms, we obtain the Hamiltonian describing the
dynamics near the resonance:
\be
\calH = -\frac{1}{2P_3^2} + \mu \,\al(P_2,P_3)\, \sin i \,  \sin(2Q_3 - Q_2 - Q_1 - \Psi),
\label{2.5}
\ee
where we introduced the notation $\al(P_2,P_3) = (a_2+b_2)/4$.

The resonance is defined by $2\dot Q_3 = \om (\tau), \, \tau = \eps t$. It follows from the
unperturbed Hamiltonian (\ref{2.3}) that $\dot Q_3 = 1/P_3^3$. Hence, denoting the value of $P_3$
at the resonance as $P_r$, we find:
\be
P_r = \left(\frac{2}{\om(\tau)} \right)^{1/3}.
\label{2.6}
\ee

Our next step is to introduce the resonant phase. We do this with the canonical transformation
$(P_i;Q_i) \mapsto (J_i;\gm_i), \, i=1,2,3$ defined with the generating function
\be
W = J_3(2Q_3 - Q_2 - Q_1 - \Psi) + J_2 Q_2 + J_1 Q_1.
\label{2.7}
\ee
For the new canonical variables we have
\be
  \begin{array}{cc}
    J_3 = P_3/2, & \gm_3 = 2Q_3 - Q_2 - Q_1 - \Psi, \\
    J_2 = P_2 + P_3/2, & \gm_2 = Q_2, \\
    J_1 = P_1 + P_3/2, & \gm_1 = Q_1.
  \end{array}
  \label{2.8}
\ee
The Hamiltonian function takes the form:
\be
\calH = -\frac{1}{8 J_3^2} + \mu \,\al(J_2,J_3)\, \sin i \,  \sin\gm_3 - \om(\tau) J_3.
\label{2.9}
\ee
Near the resonant value $P_3 = P_r$ we can expand this expression into series. With the accuracy of
order $\O(|P_3 - P_r|^3)$ we obtain the following Hamiltonian:
\be
\calF = -\frac32 \frac{(P_3 - P_r)^2}{P_r^4} + \mu \,\al(J_2,P_3)\,\sin i \,  \sin\gm_3,
\label{2.10}
\ee
where canonically conjugated variables are $P_3/2$ and $\gm_3$. Introduce notations $P=P_3, \ffi =
\gm_3 + 3\pi/2, J=J_2$. The Hamiltonian (\ref{2.10}) does not contain $\gm_2$, hence $J$ is an
integral of the problem. Another integral is $J_1$, corresponding to the fact that Delaunay element
$H$ is an integral of the original system (\ref{2.3}). Coefficient $\sin i$ in (\ref{2.10}) should
be taken at $P=P_r$. We have
\be
\left.\sin i \right|_{P=P_r} = \sqrt{1 - \frac{H^2}{(J+P_r/2)^2}}.
\label{2.10a}
\ee

From now on, we consider separately two different cases: the case of small initial eccentricity and
the case when initial eccentricity is a value of order one. Let us start with the first case.

Assume the initial value of eccentricity is small, though not necessary zero. From (\ref{2.3a}), we
have
\be
\al(J,P) \approx \frac{a e}{4} = \frac14 P\sqrt{\left( \frac{P}{2} - J\right)\left(J + \frac32
P\right)}.
\label{2.11}
\ee
Small eccentricity implies that $P/2 - J \ll 1$. As the system evolves near the resonance, small
variations of $P$ are essential when calculating the term $(P/2 - J)$ in (\ref{2.11}) and less
important in the other terms. Hence, in these latter terms, we can put $P=P_r$. We write
$$
J + \frac32 P = \left(\frac{P}{2} - J + 2J + P\right) \approx 2\left(J + \frac{P}{2}\right) \approx
2P,
$$
where we have used that $P/2\approx J$. Thus, we obtain
\be
\al(J,P)\approx\frac{\sqrt{2}}{4} P_r^{3/2}\sqrt{\frac{P}{2} - J}
\label{2.12}
\ee
and the following expression for the Hamiltonian:
\be
\calF \approx \calF_1 = -\frac32 \frac{(P - P_r)^2}{P_r^4} + \mu\sqrt{1 - \frac{H^2}{(J+P_r/2)^2}}
\, \frac{\sqrt{2}}{4} P_r^{3/2}\sqrt{\frac{P}{2} - J} \cos \ffi.
\label{2.13}
\ee
Introduce so-called Poincar\'e elements of the second kind:
\be
x = \sqrt{P-2J} \cos\ffi, \;\; y = \sqrt{P-2J} \sin\ffi.
\label{2.14}
\ee
The transformation $(P/2,\ffi) \mapsto (x,y)$ is canonical with generating function $W_1 =
\frac{y^2}{2} \cot\ffi -J\ffi$. Change the sign of $\calF_1$ and, to preserve the canonical form,
the sign of $x$. Thus, we obtain the Hamiltonian in the form:
\be
\calF_1 = \frac32 \frac{(x^2 + y^2)^2}{P_r^4} - 3\frac{x^2 + y^2}{P_r^4}(P_r - 2J) + \mu
A(H,J,P_r)x,
\label{2.15}
\ee
where
$$
A(H,J,P_r) = \frac{P_r^{3/2}}{4} \sqrt{1 - \frac{H^2}{(J+P_r/2)^2}}.
$$
Note, that if the eccentricity is small, then $P_r - 2J \ll 1$, and hence the slow variation of
$P_r$ is essential only in the second term in (\ref{2.15}). In the other terms, $P_r$ can be
assumed to be constant, say, $P_r = P_{r0}$, where $P_{r0}$ is the value of $P_r$ at $\tau = 0$.
Now, we renormalise the Hamiltonian $\calF_1 \to \frac23 P_{r0}^4 \calF_1$ to transform it to the
following standard form studied in \cite{N75}:
\be
\calF_1 = (x^2 + y^2)^2 - \la (x^2 + y^2) + \tilde \mu x
\label{2.16}
\ee
with $\la = 2(P_r - 2J)$ and $\tilde \mu = \mu A(H,J,P_{r0})$. We describe the dynamics defined by
Hamiltonian (\ref{2.16}) in Section 3.

Return now to equation (\ref{2.10}) and consider the case when the eccentricity is not small:
$e\sim 1$. In this case we can put $P = P_r$ when calculating $\al$. Thus we obtain
\be
\calF \approx \calF_2 = -\frac32 \frac{(P - P_r)^2}{P_r^4} + \mu B(H,J,P_r) \cos \ffi,
\label{2.17}
\ee
where
\be
B(H,J,P_r) = \sin i \,\al(J,P_r) = \sqrt{1 - \frac{H^2}{(J+P_r/2)^2}}\,\frac14 \left[a \calJ'_2(2e)
+ \frac{a\sqrt{1-e^2}}{e} \calJ_2(2e) \right],
\label{2.17a}
\ee
and values of $a$ and $e$ are calculated at $P=P_r$:
\be
a = P_r^2, \;\;\; e = \sqrt{1 - \frac{(J + P_r/2)^2}{P_r^2}}.
\label{2.18}
\ee
The system with Hamiltonian function (\ref{2.17}) is a pendulum with slowly changing parameters. We
study the dynamics in this system in Section 4.

\section{ Capture into the resonance at small values of eccentricity}

Dynamics in the system with Hamiltonian function (\ref{2.16}) was studied in details in \cite{N75}.
In this section we put forward the results of \cite{N75} relevant to our study.

In (\ref{2.16}) $\tilde \mu >0$ is a constant parameter, and $\la$ is a slow function of time,
$\dot\la \sim \eps$. Assume that $\dot\la >0$.

On the phase plane $(x,y)$, the values $\sqrt{P-2J}$ and $\ffi$ (see (\ref{2.14})) are polar
coordinates. Note that eccentricity in the original problem is proportional to $\sqrt{P-2J}$.

\begin{figure*}
\includegraphics{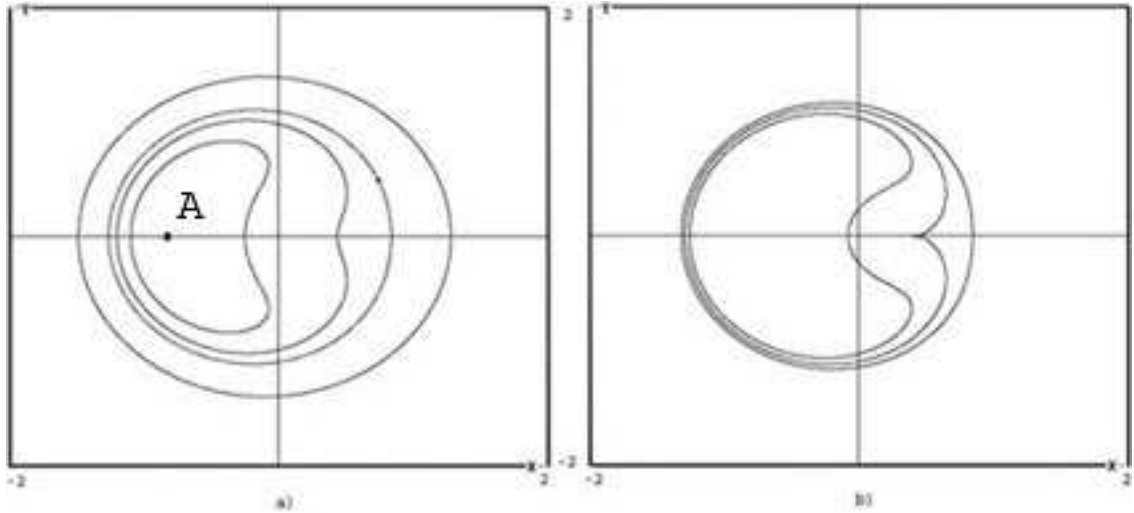}
\caption{\label{portrait12} Phase portraits of the system at fixed values of $\la$; $\tilde\mu =
0.01$. a) $\la  = 0.05 < \la_*$, b) $\la = 0.069624 \approx \la_*$.}
\end{figure*}

\begin{figure*}
\includegraphics{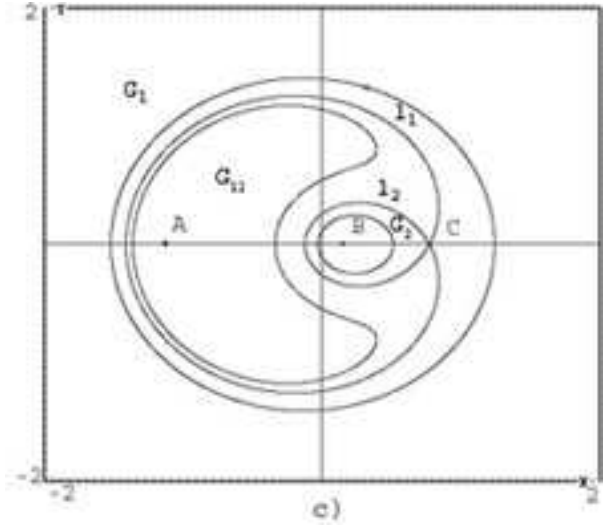}
\caption{\label{portrait3} Phase portrait of the system at $\la = 0.1 > \la_*$; $\tilde\mu =
0.01$.}
\end{figure*}

Parameter $\la$ is changing slowly, and as the first step we consider the problem at fixed values
of $\la$. Phase portraits at different values of $\la$ are presented in Figures \ref{portrait12},
\ref{portrait3}. At $\la < \la_* = \frac32 \tilde \mu^{2/3}$ (Figure \ref{portrait12}a) there is
one elliptic stable point A on the portrait. At $\la > \la_*$ (Figure \ref{portrait3}) there are
two elliptic stable points A and B, and one saddle point C. Separatrices $l_1, l_2$ divide the
phase plane into three regions $G_1, G_{12},G_2$. In Figure \ref{portrait12}b, the portrait at $\la
= \la_*$ is shown.

The coordinates of point C are $x=x_c, y = 0$, where $x_c = x_c(\la)$ is the largest root of
equation
\be
\frac{\pt\calF_1(x,0,\la)}{\pt x}  = 4x^3 - 2\la x + \tilde\mu = 0.
\label{3.1}
\ee
At $\la \ge \la_*$, introduce $F_c = F_c(\la) = \calF_1(x_c,0,\la)$ and $\tcalF (x,y,\la) = \calF
(x,y,\la) - F_c(\la)$. In $G_{12}$ we have $\tcalF < 0$, in $G_1$ and $G_2$  $\tcalF > 0$, on the
separatrices $\tcalF = 0$.

As parameter $\la$ slowly grows with time, $\dot \la \sim \eps$, curves $l_1, l_2$ defined with
$\tcalF = 0$ slowly move on the phase plane. On time intervals of order $\eps^{-1}$ their position
on the phase plane essentially changes, together with the areas of $G_{12}, G_2$. On the other
hand, area surrounded by a closed phase trajectory at a frozen value of $\la$ is an approximate
integral [adiabatic invariant] of the system with slowly varying parameter $\la$. Therefore, a
phase point can cross $l_1, l_2$ leaving one of the regions $G_i$ and entering another region.

Denote with $(x(t),y(t))$ a phase point moving according to (\ref{2.16}). Without loss of
generality assume that $\la=\la_*$ at $t=0$. The initial point $(x(0),y(0))$ can be either inside
$l_1$ [$\tcalF_0 = \tcalF (x(0),y(0),\la_*) <0$] or out of $l_1$ [$\tcalF_0 >0$]. The following
assertion is valid. All points lying inside $l_1$ at $\la=\la_*$ except, maybe, of those belonging
to a narrow strip $-k_1 \eps^{6/5} \le \tcalF_0 <0$, where $k_1$ is a positive constant, stay in
$G_{12}$ at least during time intervals of order $\eps^{-1}$.

This result is due to the fact that the area of $G_{12}$ monotonously grows with time, and
conservation of the adiabatic invariant makes a phase point go deeper and deeper into this region.
A point captured in $G_{12}$ rotates around point A$=(x_A,0)$, where $x_A$ is the smallest root of
equation (\ref{3.1}). As time grows, $|x_A|$ also grows and point A on the portrait slowly moves
along $x$-axis in the negative direction. Therefore, the motion is a composition of two components:
fast rotation along a banana-shaped curve surrounding A and slow drift along $x$-axis. The area
surrounded by each banana-shaped turn is approximately the same and equals the area $\calS (\la_*)$
surrounded by the trajectory passing through $(x(0),y(0))$ at $\la = \la_*$. Hence, the average
distance between the phase point and the origin slowly grows, corresponding to the eccentricity
growth in the original problem.

In \cite{GF}, it was shown that a point having zero initial eccentricity necessarily undergoes the
eccentricity growth. The formulated result implies that this is also valid for all the points
initially [i.e., at $\la = \la_*$] inside $l_1$, except, maybe for a narrow strip close to $l_1$. A
typical linear size of this domain is of order $\tilde\mu ^{1/3}$. In \cite{Sincl} this phenomenon
was described and called "automatic entry into libration".

Consider now the case when the point $(x(0),y(0))$ is outside $l_1$: $\tcalF_0 >0$. With time the
area inside $l_1$ grows, and at a certain moment the phase trajectory crosses $l_1$. In the
adiabatic approximation the area surrounded by the phase trajectory is constant: $\calS (\la) =
\calS (\la_*)$. Hence, in this approximation the time moment of crossing $l_1$ can be found from
equation $\calS_1 (\Lambda) = \calS (\la_*)$. Here $\calS_1 (\la)$ is the area inside $l_1$ as a
function of $\la$ and $\Lambda$ is a value of the parameter at this moment. After crossing, there
are two possibilities: (i) the phase point can continue its motion in $G_{12}$ during a time
interval of order at least $\eps^{-1}$ [this corresponds to capture into the resonance and growth
of the eccentricity]; (ii) without making a full turn inside $G_{12}$ the phase point can cross
$l_2$ and continue its motion in $G_2$ [this corresponds to passage through the resonance without
capture]. The area of $G_2$ also monotonously grows with time, hence such a point can not cross
$l_2$ once more and return to $G_{12}$

It is shown in \cite{N75} that the scenario of motion after crossing $l_1$ strongly depends on
initial conditions $(x(0),y(0))$: a small, of order $\eps$, variation of initial conditions can
result in qualitatively different evolution. If the initial conditions are defined with a final
accuracy $\dt$, $\eps \ll \dt \ll 1$, it is impossible to predict {\it a priori} the scenario of
evolution. Therefore, it is reasonable to consider capture into $G_{12}$ or $G_2$ as random events
and introduce their probabilities.

Following \cite{Arn63}, consider a circle of small radius $\dt$ with the centre at the initial
point M$_0$. Then the probability of capture into $G_{12}$ is defined as
\be
\calP = \lim_{\dt \to 0} \; \lim_{\eps \to 0} \; \frac{S_{12}}{S_{M_0}^{\dt}},
\label{3.2}
\ee
where $S_{M_0}^{\dt}$ is the measure of the circle  of radius $\dt$ and $S_{12}$ is the measure of
the points inside this circle that are captured finally into $G_{12}$.

Let $\la = \Lambda$ be the parameter value at the moment of crossing $l_1$ in the adiabatic
approximation. The following formula for probability $P$ is valid:
\be
\calP= \frac{I_1 - I_2}{I_1}, \;\; \mbox{where} \; I_1(\la) = -\oint_{l_1} \frac{\pt \tcalF}{\pt
\la} \dd t, \;\; I_2(\la) = -\oint_{l_2} \frac{\pt \tcalF}{\pt \la} \dd t,
\label{3.3}
\ee
and the integrals $I_1, I_2$ are calculated at $\la = \Lambda$. Calculating the integrals, one
finds \cite{N75}
\be
I_1(\Lambda) = \frac12 (2\pi - \Theta), \;\; I_2 = \frac{\Th}{2}, \;\; \Th = \arccos\left(
\frac{\Lambda}{2x_c^2} - 2 \right).
\label{3.4}
\ee
Here $\Th$ is the angle between the tangencies to $l_1$ at C, $0 \le \Th < \pi$.

Geometrically, formula (\ref{3.3}) can be interpreted as follows. In a Hamiltonian system, phase
volume is invariant. As parameter $\la$ changes by $\Dt \la$, a phase volume $\Dt V_{12}$ enters
the region $G_{12}$. At the same time, a volume $\Dt V_2$ leaves this region and enters $G_2$. The
relative measure of points captured in $G_{12}$ is $(\Dt V_{12} - \Dt V_2)/\Dt V_2$. The integral
$I_1$ in (\ref{3.3}) is the flow of the phase volume across $l_1$, and $I_2$ is the flow across
$l_2$. Therefore, $\calP$ gives the relative measure of points captured into $G_{12}$.

Note that, rigorously speaking, there also exists a set of initial conditions that should be
excluded from consideration. Phase trajectories with initial conditions in this set pass very close
to saddle point C, and the asymptotic results of \cite{N75} cannot be applied to them. However,
this exclusive set is small: it is shown in \cite{N75} that its relative measure is a value of
order $\sqrt{\eps}$.

\begin{figure*}
\includegraphics{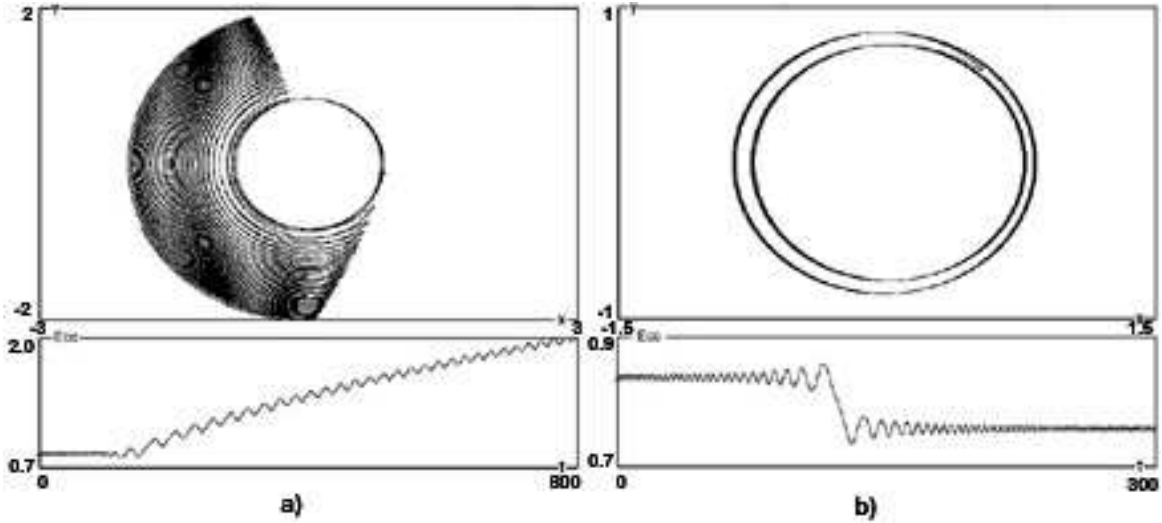}
\caption{\label{smallecc} Passage through the resonance on $(x,y)$-plane and corresponding
variation of eccentricity (bottom plots, Ecc denotes $(x^2+y^2)^{1/2}$); $\tilde\mu = 0.01, \eps =
0.01$. a) Passage with capture into the resonance; b) passage without capture; initial conditions
in the cases a) and b) are different.}
\end{figure*}

In Figure \ref{smallecc}a, capture into the resonance is shown. First, the phase point rotates
around the origin in  region $G_1$, then it crosses $l_1$, enters  region $G_{12}$ and continues
its motion in this region. In the course of this motion, the average distance from the origin
grows, corresponding to the growth of the eccentricity. In Figure \ref{smallecc}b, all the
parameter values are the same as in Figure 4a, but initial conditions are slightly different. In
this case, after crossing $l_1$, the phase point crosses $l_2$ and gets into  region $G_2$. This is
a passage through the resonance without capture.

Summarizing the results of this section, we can say the following. (i) Capture into the resonance
in the considered case results in growth of the eccentricity of the electron's orbit. (ii) On the
phase plane around the origin [$e=0$], there exists a region of size of order $\tilde\mu^{1/3}$
such that all phase trajectories with initial conditions [i.e. at $\la = \la_*$] in this region
undergo a capture into the resonance with necessity ["automatic capture"]. (iii) If initial
eccentricity is larger, and the initial point on the phase plane is out of the region mentioned
above, there is a finite probability that the phase trajectory will be captured into the resonance.
This probability is given by (\ref{3.3}).

\section{ Capture into the resonance at eccentricity of order 1.}

If initial eccentricity is a value of order one, dynamics in a $\sqrt{\mu}$-neighbourhood of the
resonance is described by Hamiltonian (\ref{2.17}). In this Hamiltonian, $P_r = P_r (\tau)$ is a
monotonously increasing function of the slow time $\tau = \eps t$.  At a frozen value of $P_r$,
this is a Hamiltonian of a pendulum, with elliptic points at $P = P_r, \, \ffi = 0 \, \mod \,2\pi$
and hyperbolic points at $P = P_r, \, \ffi = \pi \, \mod \,2\pi$. Denote the value of the
Hamiltonian at  hyperbolic points with $F_s$. The separatrices connecting the hyperbolic points
with $\ffi = -\pi$ and $\ffi = \pi$ are defined by equation $\calF_2 = F_s$ or:
\be
P-P_r = \pm \sqrt{2/3} \, P_r^2 \, \sqrt{\mu B(H,J,P_r)}\, \sqrt{1+ \cos\ffi}.
\label{4.1}
\ee
The separatrices divide the phase cylinder $P\in \mathbb{R},\, \ffi \in \mathbb{S}^1$ into domains
of direct rotation, oscillation, and reverse rotation. The area $S$ of the oscillatory domain is
proportional to $P_r^2 \sqrt{\mu B}$; it can be shown that it is a monotonically growing function
of $P_r$.

Now take into consideration the slow growth of $P_r$ with time. This growth produces slow motion of
the phase portrait upwards on the phase cylinder. At the same time the area between the
separatrices slowly grows. As a result,  phase point initially above the upper separatrix, i.e. in
the domain of direct rotation, can cross the separatrix and be either captured into the oscillatory
domain (this is a capture into the resonance), or pass through to the domain of reverse rotation.
The final mode of motion strongly depends on initial conditions: a small, of order $\eps$,
variation in them can result in qualitatively different final mode of motion. Thus, like in the
situation described in Section 3, capture into the oscillatory domain in this problem can be
considered as a probabilistic phenomenon. The probability of capture can be found as follows. As
parameter $P_r$ changes by a small value $\Dt P_r$, phase volume $\Dt V = (\frac{\dd S}{\dd P_r} +
L)\Dt P_r$, where $L$ is the length of the separatrix, crosses the upper separatrix. In the latter
expression, the first term is due to the growth of the area of the oscillatory domain, and the
second term is due to the slow motion of the upper separatrix on the phase portrait. At the same
time, phase volume $\Dt V_+ = \frac{\dd S}{\dd P_r} \Dt P_r$ enters the oscillatory domain and
stays inside of it. Hence, the probability of capture can be evaluated as
\be
\calP = \frac{\Dt V_+ }{\Dt V}.
\label{4.2}
\ee
Straightforward calculations give:
\be
\calP = \frac{\sqrt{\mu} P_r^2 K}{\sqrt{\mu} P_r^2 K + \sqrt{1+\mu P_r^6 \tilde K}},
\label{4.3}
\ee
where $K$ and $\tilde K$ are bounded functions of $P_r$ (we do not write the full expressions here
explicitly for brevity). It can be shown (see \cite{N93}) that asymptotically as $\eps \to 0$
(\ref{4.2}) gives the correct value of $\calP$ defined as in (\ref{3.2}) with $S_{12}$ denoting the
measure of captured points inside a small circle of initial conditions. At $\mu \ll P_r^{-6}$, we
find from (\ref{4.3}) that $\calP \sim \sqrt{\mu}$.

For a captured phase point value of $P$ remains at a distance of order $\sqrt{\mu}$ from $P_r$ on
time intervals of order $\eps^{-1}$. Therefore, as it follows from expressions (\ref{2.18}), as
$P_r$ grows, the eccentricity along the captured phase trajectory tends to $\sqrt{3}/2$, and the
semimajor axis of the electron's orbit tends to infinity. At $J>0$, the eccentricity is always
smaller than $\sqrt{3}/2$, and at $J<0$ it is always larger than $\sqrt{3}/2$. Note, however, that
in the original system, as it follows from (\ref{2.6}), the rate of variation of $P_r$ is large at
large values of $P_r$. Therefore, strictly speaking, the asymptotic methods used in this section
are not applicable in the limit $P_r \to \infty$.

\begin{figure*}
\includegraphics{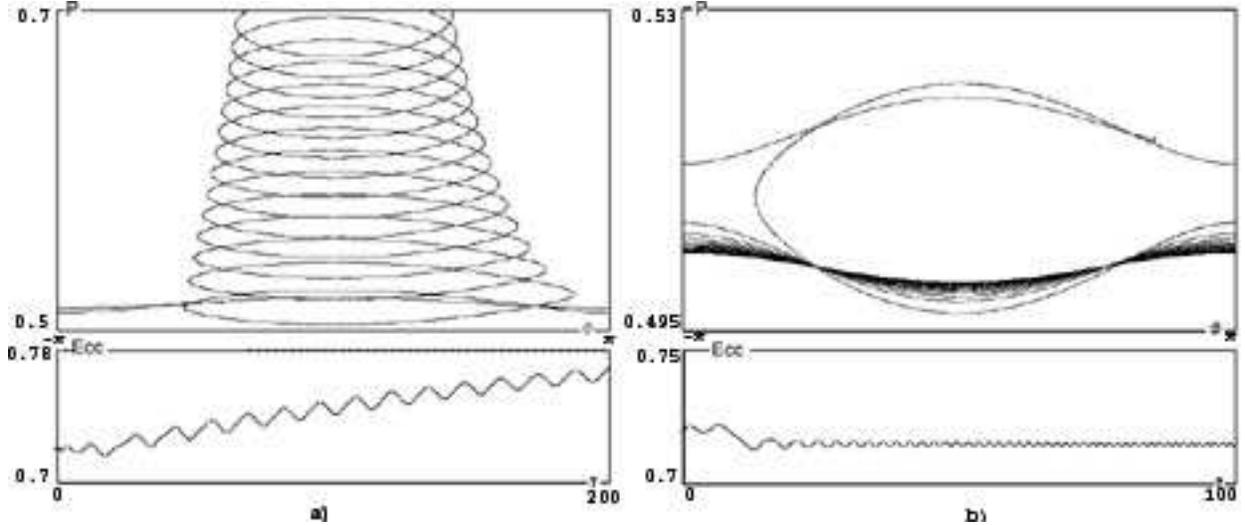}
\caption{\label{pendlike} Passage through the resonance on the $(P, \ffi)$ phase cylinder and
corresponding variation of the eccentricity (bottom plots, Ecc denotes $\sqrt{1 - \frac{(J +
P/2)^2}{P^2}}$); $\mu = 0.1, \eps = 0.001, J= 0.1, H= 0.2$. Initial conditions are different in the
cases a) and b). }
\end{figure*}

In Figure \ref{pendlike}a, capture into the resonance is shown. In the beginning, the phase
trajectory encircles the phase cylinder at approximately constant initial value of $P$. Then the
phase point crosses the upper separatrix of the pendulum (\ref{2.17}), and enters the oscillatory
domain. Since this moment, the phase trajectory does not encircle the phase cylinder and the
average value of $P$ grows. The eccentricity also grows. Figure \ref{pendlike}b shows a passage
through the resonance without capture. The phase point does not stay in the oscillatory domain, but
crosses the bottom separatrix and enters the domain of reverse rotation and continues its motion at
approximately constant new value of $P$. In this case, the eccentricity undergoes only a small
variation.

\section{Conclusions }

We have shown that if the frequency of the driving field slowly decreases, there always exists a
certain probability of capture into the resonance. A capture results in strong variation of the
electron orbit's eccentricity, and may lead to ionisation of the atom. The resonant capture
mechanism is a good tool for control of behaviour of Rydberg atoms. Note, that even if the capture
probability is small (as in the case considered in Section 4), the phenomenon is still important.
Consider, for example, an ensemble of Rydberg atoms with various initial eccentricities in the case
when the driving frequency changes slowly periodically. Then, after large enough number of these
periods, a relative number of order one of the atoms undergo the capture. If the capture
probability is a value of order $\sqrt\mu$, it will happen after $\O(\mu^{-1/2})$ periods, which
needs time of order $\eps^{-1}\mu^{-1/2}$.

\section*{Acknowledgements}

The work was partially supported with RFBR grants No. 03-01-00158 and NSch-136.2003.1 and
"Integration" grant B0053.

\end{document}